\documentclass[preprint, amsfonts, amssymb, amsmath, preprint, showkeys, floatfix, nofootinbib]{revtex4-1}
\usepackage{graphicx}
\usepackage[english]{babel}
\usepackage{float}
\usepackage[colorinlistoftodos, color=green!40, prependcaption]{todonotes}
\usepackage{amsthm}
\usepackage{tikz}
\usepackage{capt-of}
\usepackage{ bbold }
\usepackage{amsmath}
\usepackage{mathrsfs}
\usepackage{upgreek}
\usepackage{slashed}
\usepackage{mathtools}
\usepackage{physics}
\usepackage{simpler-wick}
\usepackage{tabularx}
\usepackage{xcolor}
\usepackage{graphicx}
\usepackage{svg}
\usepackage{adjustbox}
\usepackage{enumerate}
\usepackage{dsfont}
\usepackage{placeins}
\usepackage{fancyhdr}
\usepackage[T1]{fontenc}
\usepackage{lipsum}
\usepackage{csquotes}
\usepackage[normalem]{ulem}
\usepackage[pdftex, pdftitle={Article}, pdfauthor={Author}]{hyperref} 
\usepackage[normalem]{ulem}

\begin{document}
\title{Clustering in hadrons and light nuclei from Lorentz boosted form factors}

\author{F.E. Rodríguez Barrera}
\email{fe.rodriguezb1@uniandes.edu.co}
\author{N.G. Kelkar}
\email{nkelkar@uniandes.edu.co}

\affiliation{Universidad de los Andes, Bogotá, Colombia.}

\date{\today} 

\begin{abstract}
The determination of nuclear charge radii is crucial for understanding the internal structure of nuclei and their fundamental interactions. A persistent discrepancy, not only in the measured proton charge radius but also in the light nucleus charge radius, between electron scattering and muonic spectroscopy has fueled ongoing debates in nuclear and particle physics. Using this discrepancy, we revisit the role of one of the proposed solutions, namely the use of Lorentz-boosted nuclear form factors to find a subtle connection between the boost and the cluster structure of nuclei. By applying two distinct relativistic formalisms, namely the Licht-Pagnamenta and Mitra-Kumari approaches, we systematically analyze corrections to the moments of the density distributions in hadrons and nuclei. Our results demonstrate that boosting the form factors from the Breit to the rest frame of the nucleus not only assists in reconciling the spectroscopic and scattering measurements but also provides a method to infer on the quark and nucleon cluster configurations within nuclei.

\end{abstract}

\maketitle


\section{Introduction}
The size of the proton has been a topic of intense scientific inquiry for nearly a century, yet it remains one of the most puzzling and debated aspects of particle physics. As a fundamental building block of matter, the proton plays a crucial role in the structure of atomic nuclei, which drives our interest in understanding the underlying theory behind it.
The usual description using Quantum Electrodynamics involves the use of form factors (FF), a well-understood tool that allows us to describe interactions between complex particles \cite{Marek,gao,Sachs62}. The success in the use of FF is wide and has enlightened some mysteries related to time-reversal symmetry breaking processes \cite{dipoleff} or the corrections for the values of the finite size effects on atomic and particle physics experiments \cite{fsetheory, Corrected_cs, Corrected_decay, Corrected_el}. In the non-relativistic limit, the Fourier transforms of the electric and magnetic form factors, $G_{E,M}(Q^2)$, in the Breit frame (a special frame where the energy transfer is zero so that 
$q^2 = \omega^2 - {\bf q}^2 = -{\bf q}^2 = Q^2$) can be used to obtain 
the charge and magnetization density distributions inside a nucleus (or nucleon). The derivative of these form factors with respect to $Q^2$ at $Q^2 =0$ can be used to deduce the charge and magnetic radii from scattering experiments.
Although electron-hadron scattering has been used to measure the radii since the mid-1950s, it is well known that fulfilling the condition $Q^2=0$ presents significant experimental challenges. Traditional methods involve using parametrizations of $G_E(Q^2)$ to extrapolate data in the low-momentum regime, though these methods are highly dependent on the specific functions chosen \cite{Sick_2018}. This limitation can be overcome by employing a more statistical approach like the Padé type analytical continuation, which offer a more robust and flexible framework for addressing the problem. These methods reduce the reliance on specific parametrizations and provide more accurate extrapolations in the low-momentum regime \cite{Stat1,Stat2}. 

In 2010, the Committee on Data of the International Science Council (CODATA) published their periodical report on the suggested values for the common physical constants, nonetheless, the subsection of the nuclear charge radius of the proton included two different values several standard deviations away. In particular, for $ep$ scattering $r_p=0.8768(79)fm$ and for muonic spectroscopy $r_p=0.84184(67)fm$  \cite{codata2010}. This discrepancy started a new wave of experiments, using both methods, and theories, including new `Beyond the Standard Model' formalisms, to solve the `proton radius puzzle' \cite{Exp1,Exp2,Exp3,Exp4,Exp5}.The situation worsened in 2016 when a similar issue occurred with deuterium, raising concerns about the consistency of the measurements. A few years later, this pattern repeated with helium-3 and helium-4, further complicating the interpretation of experimental data. 
In \cite{Miller2019}, the author showed that a bound proton is larger than a free one. 
A combined analysis of the electromagnetic form factors of the nucleon in space- and time-like regions using dispersion theory is worth mentioning \cite{LinHammer2022, Lorenz2012}. The proton charge radius, $r_p = 0.840^{+0.003 + 0.002}_{-0.002-0.002}$ fm, extracted by the authors, is close to the spectroscopic value. 

Since the discrepancy arises from the use of two different experimental methods, it is essential to
recognize the differences between them. Scattering experiments are typically conducted in high-energy and high-momentum-transfer regimes, whereas atomic spectroscopy is performed in a rest frame. Therefore, it is natural to consider applying a Lorentz boost to relate the two frames and account for relativistic corrections to the measurements. This approach is further supported by the fact that the radii of nuclei are essentially lengths; even though the definition of electromagnetic form factors (FF) originates from a relativistic framework, the length of a nucleus should not remain invariant. After applying the Lorentz boost and incorporating a QED correction from the Breit equation, the proton radius exhibits a shift of approximately 4\% relative to the value obtained from electron scattering experiments \cite{neelimalcp}. This idea is also consistent with the results of the PRad experiment on low momentum transfer with a proton radius value closer to the spectroscopy-related value \cite{PRAD}. 

There exists a great deal of literature \cite{jentschura2022, lumpay2025, qi2025} that provides models and theories to explain the discrepancy between the spectroscopic and scattering measurements of the radii (see \cite{HillEPJweb2017,Gao2022} for a review). {\it It is not the objective of the present work to enter into this debate, but rather to use the discrepancy as an advantage to comment on the clustering of the constituents in a nucleus}. With this objective, we investigate the 
available formalisms for the Lorentz boost of form factors.
One of the first proposals of applying a Lorentz boost to rewrite form factors in 
the rest frame in terms of those in a moving frame was put forth by Licht and Pagnamenta (LP) \cite{LichtPagnamenta1}. Modifications and generalizations
of this proposal followed later (as will be seen in the next section). 
An interesting feature of the LP formalism was that the relation between form factors in two
different frames depended on the number of constituents that made up the object. The latter helps to address the fundamental question whether a nucleus is simply a bound state of nucleons or clusters of light nuclei. 

The article is organized as follows: in the next section we shall briefly review the formalisms presented by Licht and Pagnamenta \cite{LichtPagnamenta1} and Mitra and Kumari \cite{MitraKumari} for the 
Lorentz boost of form factors. Based on these formalisms, we shall derive a relation between the radii in a moving frame and that at rest. The conversion is found to depend on the number, $n$, of the clusters inside the object. For example, $n$ = 3 for a proton with 3 quarks,  
$n$ = 2 for a deuteron with 2 nucleons or $n$ = 6 for a deuteron with six quarks.  In section \ref{results}, using the available values of radii 
from scattering and spectroscopic experiments, we find the most suitable $n$ to satisfy the relation between radii in the moving and rest frame.
Based on the values of $n$, we comment on the possible clustering 
in hadrons and nuclei. We caution the reader that the present work is not an attempt to comment on the possible clustering in hadrons or nuclei based on fundamental theories of the strong interaction but rather to use a subtle connection between the radii from scattering and spectroscopic measurements which depends on the number of constituent clusters. 
The results seem to be consistent with expectations from fundamental theories and literature.
Section \ref{summary} provides a summary of the 
present work.

\section{Electromagnetic form factors and radii}\label{formalisms}
Let us begin by understanding the need for a Lorentz boost of the form factors. Composite systems of quarks and nucleons can usually be well described by non-relativistic wave functions and form factors in their rest frames. However, to extract information on their size, one may resort to high-energy scattering experiments. The second moment of the 
spatial density distribution of the constituents in a composite system, 
$\rho(\bf{r})$, commonly referred to as the radius of the system, 
can be given as
\begin{equation}
    \langle r^2 \rangle = \int \,r^2 \, \rho({\bf r}) \, d^3r
\end{equation}
where, the density $\rho({\bf r})$ can be related to its momentum space 
counter part through a Fourier transform. The momentum space form factors 
are determined from scattering experiments and the standard way of defining the $n^{th}$ moment of a charge or magnetization distribution 
involves the Fourier transform of the so-called Sachs form factors in 
the Breit frame (defined by $\omega = 0$, with the four momentum squared
$q^2 = \omega^2 - {\bf q}^2$). For example, starting with, 
$G({\bf q}^2) = \int \, e^{-i{\bf q} \dot {\bf r}} \rho({\bf r}) d^3r / (2\pi)^3$, one can easily show that
\begin{equation}
    G({\bf q^2}) = {1 \over 2\pi^2} \,\biggl [  \int_0^{\infty} r^2 \rho(r) dr - {{\bf q}^2 \over 6} \int_0^{\infty} r^4 \rho(r) dr + ... 
    \biggr ]\,,
\end{equation}
leading to the standard definition 
\begin{equation}\label{radiusdef}
    {-6 \over G(0)} \, {dG({\bf q}^2) \over d{\bf q}^2}\biggl|_{{\bf q}^2 =0} \, =\, \int r^4 \rho(r) dr = \langle r^2\rangle\,. 
\end{equation}
Note that even if one starts with the Lorentz invariant form factor 
$G(q^2)$, by taking the derivative at $q^2 =0$, in the Breit frame with 
$\omega =0$, one arrives at the definition of the radius, which is a length and hence not a Lorentz invariant quantity. Hence, if one wishes 
to compare the radius obtained from a scattering experiment with that determined using spectroscopic methods where the nucleus is at rest, it 
is necessary to perform a Lorentz boost from the Breit frame to the rest 
frame of the nucleus.
\subsection{Relativistic clusters of Licht and Pagnamenta}
The first attempt to apply a Lorentz boost to wave functions and form factors of composite particles was made in the 70's when parton and quark models started arising and were a subject of debate. Noting that wave functions in the rest frame and non-relativistic form factors are often used in the calculations of composite objects, 
Licht and Pagnamenta (also referred to as LP) showed how the wave functions should be 
boosted to relativistic velocities and how this affects the form factors which enter 
the differential cross sections.
LP proposed a simple model for a spinless composite object in elastic scattering in which the relativistic form factor $G(q^2)$ is given as \cite{LichtPagnamenta1}, 
\begin{equation}\label{LPformfactors}
    G(q^2)=\left( 1-\frac{q^2}{4M_A^2}  \right)^{(1-n)/2}G^0\left(|q^2|/\left(1-\frac{q^2}{4M_A^2}\right)\right),
\end{equation}
where $G^0$ is the non-relativistic form factor, n is the number of particles in the cluster and $M_A$ is the total mass of the cluster. 
Thus, in order to obtain a relation between the radius in the rest frame of the nucleus and that from a scattering experiment, we associate 
$G(q^2)$ with the Sachs form factor in the Breit frame and $G^0$ to that 
in the rest frame of the nucleus. Taking the derivative of the form factors with respect to $q^2$ and using the definition in (\ref{radiusdef}) for the radius, after some algebra, we get    
\begin{equation} \label{nice}
    <r^2>= <r^2>^0+ \frac{3(n-1)}{4M_A^2}
\end{equation}
where $<r^2>^0$ is the radius in the rest frame of the nucleus. This relation can be used to obtain an estimate of any nuclear charge or magnetic radius depending on the number of particles, $n$, in the cluster.

\subsection{Generalization of the LP argument}
Few years after the first attempt to apply the Lorentz boost, a modified version of the Licht and Pagnamenta's argument was made by Mitra and Kumari 
\cite{MitraKumari} in order to improve some of the aspects of the original formula. These corrections had the purpose to include a symmetrical treatment between the initial and final states and to have a better asymptotic behaviour in the large momentum regime. The proposed formula 
is \cite{MitraKumari}
\begin{equation}
    G_{AB}(q^2)=\left( \frac{M_AM_B}{E_A E_B} \right)^{n-1}G^0_{AB}(4p^2\gamma_p^2)
\end{equation}
where $M_{A,B}$ and $E_{A,B}$ refer to the mass and energy of the initial A or final B state, $\gamma_p \coloneqq (M_AE_A^{-1} + M_BE_B^{-1})/2$ and $p$ is the Breit-frame momentum. Note that the explicit dependence on the initial and final states broadens the use of the Lorentz transformation to inelastic scattering. Since we are interested in elastic scattering, the masses are equal and $p=q/2$ so that the above formula becomes
\begin{equation}
    G(q^2)=\left( 1-\frac{q^2}{4M_A^2} \right)^{1-n} G^0 \left( \frac{q^2}{1-\frac{q^2}{4M_A^2}} \right),
\end{equation}
which is quite similar to the previous formula (\ref{LPformfactors}) by LP, with the difference in the exponent of the correction term. This difference is notable since as mentioned before, it gives the correct asymptotic behaviour of the form factor as predicted by QCD in terms of the number of constituents \cite{asymptotic}. Computing the radius once again as above, the correction has the form 
\begin{equation}\label{MKradius}
    <r^2>=<r^2>^0+\frac{3(n-1)}{2M_A^2}.
\end{equation}

Even if this formula is fairly similar to the original correction, this refined approach not only extends the applicability of Lorentz transformations to inelastic processes but also enhances the theoretical consistency of the form factors in the high-momentum regime. 

\section{Number of clusters and radii}\label{results}
Naively speaking, a nucleus is made up of protons and neutrons (nucleons) and hadrons are composites of quarks and antiquarks. However, some nucleons inside a nucleus may group together and exhibit a collective 
behavior \cite{WeiNST2024}. The $\alpha$ decay of heavy nuclei gave us one of the first clues to such possibilities. However, clustering is not restricted to 
heavy unstable nuclei. The 7.65 MeV state in $^{12}$C, also known as the Hoyle state \cite{HoyleApJS1954}, plays an important role in nucleosynthesis and has been established to be a 3 $\alpha$ cluster. At 7.16 MeV, $^{16}$O exhibits a 
$^{12}$C + $\alpha$ substructure and a 4 $\alpha$ cluster structure at 
14.44 MeV \cite{FreerRPP2007}. The often cited $^{16}$O + $\alpha$ cluster structure of 
$^{20}$Ne is confirmed by experiment \cite{BuckPRC1995}. 
An inelastic $\alpha$-scattering experiment on the unstable N = Z, 
doubly-magic $^{56}$Ni nucleus performed in inverse kinematics at an incident energy of 50 A.MeV at GANIL \cite{BagchiEPJA2020} provided 
signature of a multiple $\alpha$ (up to 14) structure of $^{56}$Ni. 

In spite of the above, one cannot forget that the nucleus is bound by the strong interaction and the fundamental theory of the strong interaction is Quantum Chromodynamics (QCD), the theory of quarks and gluons with charges called ``color quantum numbers". The spectroscopy of hadrons has played a major role in the development of QCD \cite{FranzGrossEPJC2023} where baryons are typically described by a $qqq$ and mesons by a $q \bar{q}$ configuration. It is then natural to probe into the possibility of nuclei displaying not just a multi-nucleon or multi-cluster structure but rather a multi-quark \cite{tsaiPTP1980} one with zero overall color. The
existence of six-quark cluster components in nuclei and its 
implications have been discussed in literature \cite{millerPRL1984, kochmillerPRC1985,kankiPTP1985} with the lightest nucleus, the deuteron,  being the most probable example \cite{glozmanPPNP1995,burovZPA1982}. 
In what follows, we shall present results on the comparison of hadronic and nuclear radii from scattering and spectroscopy (Eqs (\ref{nice}) and (\ref{MKradius})) using different values, 
$n$, for the number of clusters inside the hadrons and nuclei.
\subsection{Proton charge and magnetic radius}\label{proton}
Starting with the corrections to the charge radius of the proton, we use the most recent suggested value from CODATA using $e-p$ scattering, which is less precise than the previous values but incorporates the data up to 2020. For the muonic spectroscopy experimental data we use the value obtained in 2010 that started the proton radius puzzle discussion. By considering the proton as a cluster made up of 3 quarks ($uud$) the corrections for the radius can be easily calculated.

\def\cellwidth{40}
\def\cellheight{7}
\def\whitecellcolor{white}

\begin{figure}[H]
    \centering
    \begin{tikzpicture}[%
      x=1mm,%
    y=-1mm,%
    every node/.style={%
        shape=rectangle,
        anchor=north west,
        outer sep=0mm,
        draw=black},
    whitecell/.style={%
        fill=\whitecellcolor,
        minimum height=\cellheight mm,
        minimum width=\cellwidth mm},
    bigwhitecell/.style={%
        fill=\whitecellcolor,
        minimum height=\cellheight mm,
        minimum width=120 mm}]

    \node[whitecell] at (0,0){Formalism};
    \node[whitecell] at (40,0){Charge Radius (fm)};
    \node[whitecell] at (80,0){Magnetic radius (fm)};
    \node[whitecell] at (0,7){Licht-Pagnamenta};
    \node[whitecell] at (40,7){\textbf{0.84157}};
    \node[whitecell] at (80,7){0.810};
    \node[whitecell] at (0,14){Mitra-Kumari};
    \node[whitecell] at (40,14){0.80129};
    \node[whitecell] at (80,14){\textbf{0.768}};
    \node[bigwhitecell] at (0,21){Experimental Reference values (fm)};
    \node[whitecell] at (0,28){Scattering};
    \node[whitecell] at (40,28){0.880(20)};
    \node[whitecell] at (80,28){0.850(10)};
    \node[whitecell] at (0,35){Spectroscopy};
    \node[whitecell] at (40,35){0.84184(67)};
    \node[whitecell] at (80,35){0.778(29)};

    \end{tikzpicture}

    \captionof{table}{Corrections to the proton charge $\sqrt{<r^2>^0_E}$ and magnetic radius $\sqrt{<r^2>^0_M}$ in fm for every formalism. The scattering reference value for both radii is an average from the world data on $e-p$ scattering experiments \cite{codata2018, magneticprot} (measured $\sqrt{<r^2>}$) and the spectroscopy value is obtained from $\mu H$ experiments \cite{CREMAH} for charge radius and hyperfine splitting in H \cite{truemagprot} (measured $\sqrt{<r^2>^0}$).The numbers in boldface indicate the Lorentz boosted radius which agrees best
    with the spectroscopy value.}
    \label{table:proton}
\end{figure}

As can be seen in Table \ref{table:proton}, the correction shrinks the value of the 
radius deduced from scattering experiments with both the formalisms used and brings it closer to the value from spectroscopy. The Mitra-Kumari formalism leads to an underestimation of the proton's radius that may be partially solved by considering more corrections in the calculations. For the magnetic radius most of the values are extracted using $e-p$ scattering using the same methods for the charge radius. However, using atomic methods for this radius is far more complex. One must use the Zemach contribution to the hyperfine splitting and do a fit to find the form factor. Since the fits of form factors have some issues as pointed out in the introduction, some attempts of finding a model independent method has been proposed \cite{specmagprot}. In this case, the Mitra-Kumari formalism gives a better approximation to the reported measurements.
\subsection{Deuteron charge radius}\label{deuteron}
The example of the proton charge and magnetic radii showed us that assuming the proton 
to be a cluster of 3 quarks, we could indeed reduce the discrepancy between the scattering 
and spectroscopic data in the right direction. Hence, we now propose to use the 
discrepancy in the scattering and spectroscopic data and vary $n$ in Eqs (\ref{nice}) and 
(\ref{MKradius}), treating it as a parameter to find the value which best reduces 
the discrepancy. The choice of the values of $n$ is based on the nucleus under consideration.
For the deuteron, we shall consider $n$ = 2 and $n$ = 6 as two possibilities that the deuteron is a composite of two nucleons or six quarks.

In case of the deuteron, a discrepancy between experimental values emerged after the CREMA collaboration published their results for muonic deuterium. Given the results of Table \ref{table:deuterium}, the best value is obtained by using the Mitra-Kumari approach and $n=6$. This result is certainly interesting considering that the deuteron's wave function traditionally has been obtained by considering this nucleus as being made up of two nucleons. However, there are some experimental signals of the six-quark clustering in deuteron \cite{6quarkexp}. 
The result does not necessarily hint towards the deuteron being a hexaquark state but rather a pair of baryons composed of 3 quarks each. There do exist theoretical models of deuterium as a 6 quark singlet state \cite{6quarks}, though.

\begin{figure}[H]
    \centering
    \begin{tikzpicture}[%
      x=1mm,%
    y=-1mm,%
    every node/.style={%
        shape=rectangle,
        anchor=north west,
        outer sep=0mm,
        draw=black},
    whitecell/.style={%
        fill=\whitecellcolor,
        minimum height=\cellheight mm,
        minimum width=\cellwidth mm},
    medwhitecell/.style={%
        fill=\whitecellcolor,
        minimum height=\cellheight mm,
        minimum width=80 mm},
    bigwhitecell/.style={%
        fill=\whitecellcolor,
        minimum height=\cellheight mm,
        minimum width=120 mm}]

    \node[whitecell] at (0,0){Formalism};
    \node[whitecell] at (40,0){$n=2$ (fm)};
    \node[whitecell] at (80,0){$n=6$ (fm)};
    \node[whitecell] at (0,7){Licht-Pagnamenta};
    \node[whitecell] at (40,7){2.13943};
    \node[whitecell] at (80,7){2.13163};
    \node[whitecell] at (0,14){Mitra-Kumari};
    \node[whitecell] at (40,14){2.13744};
    \node[whitecell] at (80,14){\textbf{2.12192}};
    \node[bigwhitecell] at (0,21){Experimental Reference values (fm)};
    \node[whitecell] at (0,28){Scattering};
    \node[medwhitecell] at (40,28){2.1413(25)};
    \node[whitecell] at (0,35){Spectroscopy};
    \node[medwhitecell] at (40,35){2.12562(13)};

    \end{tikzpicture}

    \captionof{table}{Corrections to the deuteron charge radius in fm for different formalisms. The scattering reference value is an average from the world data \cite{codata2014} (measured $\sqrt{<r^2>}$) given by CODATA and the spectroscopy value is obtained from $\mu D$ experiments \cite{CREMAD}(measured $\sqrt{<r^2>^0}$) done by the CREMA collaboration.The bold faced value indicates the Lorentz boosted radius which agrees best
    with the spectroscopy value.}
    \label{table:deuterium}
\end{figure}
\subsection{Helium isotopes $^3$He and $^4$He and the triton}\label{helium}
The next element of interest is Helium, with Helium-3 being the lightest stable isotope. The possibilities of constituents can be 3 nucleons or 9 quarks.
The possibility of six- and nine-quark interior states was considered in 
\cite{maizePRC1985}. With the bound-state $^3$He wave function, 
which has the interior multi-quark state confined within a certain 
cutoff radius and an exterior three-nucleon state, the authors calculated
the He charge form factor using the relativistic harmonic oscillator quark model to find a small contribution, namely, 2.7\% of the six- and 
0.03\% of the nine-quark interior states. 
The results in Table \ref{table:he3} show that the 3 nucleon configuration gives better results in both formalisms.  
Taking the larger $n$, results in a bigger underestimation of the charge radius so it seems that from here on the general behaviour of nuclei will be determined by the number of nucleons and not the number of valence quarks. This result is also consistent with the 
small multi-quark contribution found in \cite{maizePRC1985}. 

\begin{figure}[H]
    \centering
    \begin{tikzpicture}[%
      x=1mm,%
    y=-1mm,%
    every node/.style={%
        shape=rectangle,
        anchor=north west,
        outer sep=0mm,
        draw=black},
    whitecell/.style={%
        fill=\whitecellcolor,
        minimum height=\cellheight mm,
        minimum width=\cellwidth mm},
    specialcell/.style={%
        fill=unmellowyellow,
        minimum height=\cellheight mm,
        minimum width=\cellwidth mm},
    medwhitecell/.style={%
        fill=\whitecellcolor,
        minimum height=\cellheight mm,
        minimum width=80 mm},
    bigwhitecell/.style={%
        fill=\whitecellcolor,
        minimum height=\cellheight mm,
        minimum width=120 mm}]

    \node[whitecell] at (0,0){Formalism};
    \node[whitecell] at (40,0){$n=3$ (fm)};
    \node[whitecell] at (80,0){$n=9$ (fm)};
    \node[whitecell] at (0,7){Licht-Pagnamenta};
    \node[whitecell] at (40,7){1.97114};
    \node[whitecell] at (80,7){1.96554};
    \node[whitecell] at (0,14){Mitra-Kumari};
    \node[whitecell] at (40,14){\textbf{1.96927}};
    \node[whitecell] at (80,14){1.95805};
    \node[bigwhitecell] at (0,21){Experimental Reference values (fm)};
    \node[whitecell] at (0,28){Scattering};
    \node[medwhitecell] at (40,28){1.973(14)};
    \node[whitecell] at (0,35){Spectroscopy};
    \node[medwhitecell] at (40,35){1.97007(94)};

    \end{tikzpicture}

    \captionof{table}{Corrections to the Helium-3 nuclear charge radius in fm for different formalisms. The scattering reference value is an average from the world data experiments of electron elastic scattering \cite{he3scattering} (measured $\sqrt{<r^2>}$) given by Ingo Sick and the spectroscopy value is obtained from $\mu \prescript{3}{}{He}$ experiments \cite{he3muon} (measured $\sqrt{<r^2>^0}$) done by the CREMA collaboration.The bold faced value indicates the Lorentz boosted radius which agrees best with the spectroscopy value.}
    \label{table:he3}
\end{figure}

There exists data on the charge radius of $^3$He from scattering as well a 
precise value from spectroscopy, however, we do not have a precise spectroscopic value for 
$^3$H. $^3$He and $^3$H are the simplest pair of mirror nuclei, namely, nuclei with the same number of nucleons but the number of protons and neutrons swapped. The $^3$He charge radius of 1.973(14) fm \cite{he3scattering} is larger than the charge radius 1.7591(363) fm \cite{angeli2013} of $^3$H. Naively, one would say that the Coulomb repulsion in 
$^3$He is larger than in $^3$H 
and could be one of the reasons for the difference. 
The Lorentz boost formalism presented here seeks to use the discrepancy between the 
scattering and spectroscopic values of radii (in different reference frames) of a given nucleus to investigate the cluster structure of that nucleus. It cannot be used to explain the difference in the radii of mirror nuclei. 
The difference in the charge radii of mirror nuclei is related to the difference in the nuclear (and Coulomb) forces and we refer the reader to some interesting articles such as \cite{jaffe1968} (and references therein) where the authors used separable nucleon-nucleon potentials and solved Faddeev equations, Ref. \cite{folk1968} and a more recent calculation of the triton charge radius using next-to-next-to-leading order pionless effective field theory \cite{vanasse2017}.  On the experimental side, 
a Jefferson lab experiment 
\cite{Myers2016} proposed to determine the ratio of the electric form factors of these $A$ = 3 nuclei. More recently, in Mainz, the T-REX experiment \cite{pohltalk} is being set up to determine the triton charge radius by laser spectroscopy of atomic tritium. In 
Table \ref{triton} we give our prediction of the possible spectroscopic value of the $^3$H charge radius after applying the correction to the experimental radius from scattering. 
\begin{figure}[H]
    \centering
    \begin{tikzpicture}[%
      x=1mm,%
    y=-1mm,%
    every node/.style={%
        shape=rectangle,
        anchor=north west,
        outer sep=0mm,
        draw=black},
    whitecell/.style={%
        fill=\whitecellcolor,
        minimum height=\cellheight mm,
        minimum width=\cellwidth mm},
    specialcell/.style={%
        fill=unmellowyellow,
        minimum height=\cellheight mm,
        minimum width=\cellwidth mm},
    medwhitecell/.style={%
        fill=\whitecellcolor,
        minimum height=\cellheight mm,
        minimum width=80 mm},
    bigwhitecell/.style={%
        fill=\whitecellcolor,
        minimum height=\cellheight mm,
        minimum width=120 mm}]

    \node[whitecell] at (0,0){Formalism};
    \node[whitecell] at (40,0){$n=3$ (fm)};
    \node[whitecell] at (80,0){$n=9$ (fm)};
    \node[whitecell] at (0,7){Licht-Pagnamenta};
    \node[whitecell] at (40,7){1.75700};
    \node[whitecell] at (80,7){1.75067};
    \node[whitecell] at (0,14){Mitra-Kumari};
    \node[whitecell] at (40,14){{1.75489}};
    \node[whitecell] at (80,14){1.74220};
    \node[bigwhitecell] at (0,21){Experimental Reference values (fm)};
    \node[whitecell] at (0,28){Scattering};
    \node[medwhitecell] at (40,28){1.7591(363)};
    \node[whitecell] at (0,35){Spectroscopy};
    \node[medwhitecell] at (40,35){ T-REX in future \cite{pohltalk}};

    \end{tikzpicture}

    \captionof{table}{Corrections to the triton nuclear charge radius in fm for different formalisms. The scattering reference value is from \cite{angeli2013}.}
    \label{triton}
\end{figure}

For the Helium-4 isotope we have many possibilities associated to its internal structure. The most reasonable possibility is that the $^4$He nucleus behaves as 4 nucleons and not 12 quarks or a bound state of 2 deuterons. This is reinforced by the results in Table \ref{table:he4}, where in both formalisms the best result comes from $n=4$. Most nuclear models do not use a mix of 12 quark wave functions but instead only 4 nucleons. Also, considering 2 deuterons would not be physically appealing due to the stability of Helium-4 which is also the reason behind the fact that the fusion of 2 deuterons will typically give tritium or Helium-3. Now, even if the CREMA collaboration gives values for the other isotopes of Helium via isotope shift, the scattering of $^{6,8}$He is mainly used to obtain results for the matter distribution radius and not the charge radius. Both of these nuclei are treated as a Helium-4 nucleus surrounded by a halo of neutrons, hence increasing the apparent radius obtained from scattering data.

\def\cellwidth{40}
\begin{figure}[H]
    \centering
    \begin{tikzpicture}[%
      x=1mm,%
    y=-1mm,%
    every node/.style={%
        shape=rectangle,
        anchor=north west,
        outer sep=0mm,
        draw=black},
    whitecell/.style={%
        fill=\whitecellcolor,
        minimum height=\cellheight mm,
        minimum width=\cellwidth mm},
    medwhitecell/.style={%
        fill=\whitecellcolor,
        minimum height=\cellheight mm,
        minimum width=120 mm},
    bigwhitecell/.style={%
        fill=\whitecellcolor,
        minimum height=\cellheight mm,
        minimum width=160 mm}]

    \node[whitecell] at (0,0){Formalism};
    \node[whitecell] at (40,0){$n=2$ (fm)};
    \node[whitecell] at (80,0){$n=4$ (fm)};
    \node[whitecell] at (120,0){$n=12$ (fm)};
    \node[whitecell] at (0,7){Licht-Pagnamenta};
    \node[whitecell] at (40,7){1.68038};
    \node[whitecell] at (80,7){\textbf{1.67916}};
    \node[whitecell] at (120,7){1.67423};
    \node[whitecell] at (0,14){Mitra-Kumari};
    \node[whitecell] at (40,14){1.67977};
    \node[whitecell] at (80,14){1.67730};
    \node[whitecell] at (120,14){1.66743};
    \node[bigwhitecell] at (0,21){Experimental Reference values (fm)};
    \node[whitecell] at (0,28){Scattering};
    \node[medwhitecell] at (40,28){1.6810(40)};
    \node[whitecell] at (0,35){Spectroscopy};
    \node[medwhitecell] at (40,35){1.67824(13)};

    \end{tikzpicture}

    \captionof{table}{Corrections to the Helium-4 nuclear charge radius in fm for different formalisms. The scattering reference value is an average from the world data experiments of electron and proton scattering \cite{he4scattering} (measured $\sqrt{<r^2>}$) given by Ingo Sick and the spectroscopy value is obtained from $\mu \prescript{4}{}{He}$ experiments \cite{he4muon} (measured $\sqrt{<r^2>^0}$) done by the CREMA collaboration.The bold faced value indicates the Lorentz boosted radius which agrees best
    with the spectroscopy value.}
    \label{table:he4}
\end{figure}

The fact that the Helium-4 isotope gives a very good correction with a value of $n=4$ allows us to propose a novel method for studying the precluster or subcluster structure of any nucleus. By taking precise enough measurements of the nuclear charge radius using scattering and comparing the value with the one obtained from muonic spectroscopy, the best fit of $n$ can reveal the existence of internal clusters. For example, one may expect a value of $n=4$ in the case of $^{16}$O or $n=3$ in the case of $^{12}$C due to phenomena like the $\alpha$-particle condensation that has been studied in both theoretical and experimental ways for these nuclei \cite{16o1,16o2,16o3}. The existence of these states has been widely researched and used to study and explain nuclear properties or events, like the Gamow theory of alpha decay or, in general, the tendency of some nuclei to decay into specific and more stable states.

\begin{figure}[H]
    \centering
    \includegraphics[scale=0.5]{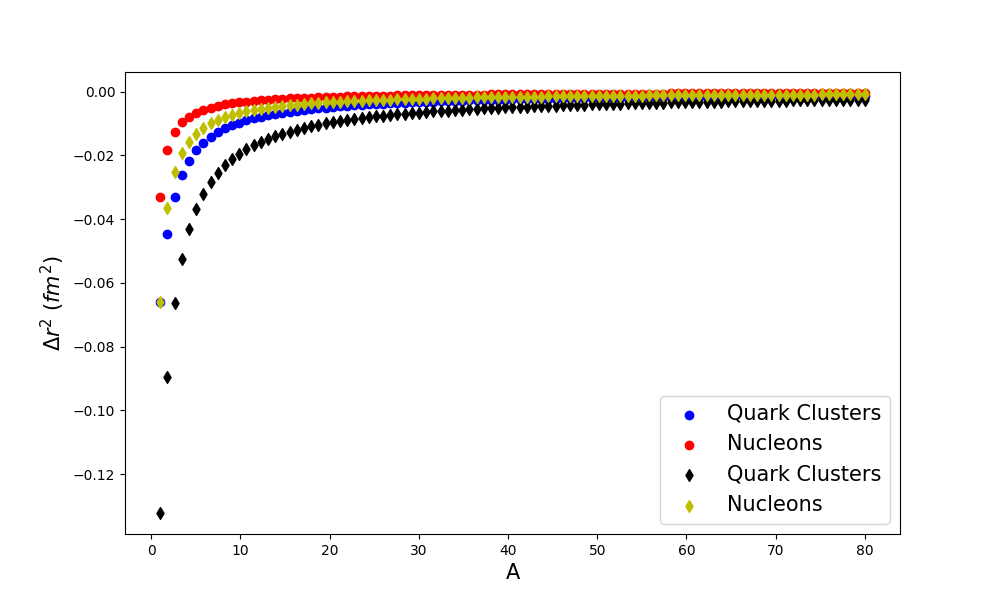}
    \caption{Graph of the expected correction for several nuclei as a function of the number of nucleons A, considering two scenarios: (i) the 
    nucleus is made up entirely of quarks with the number of quarks being simply 3A and (ii) the nucleus is made of A nucleons.
    Filled circles display the calculations in the Licht-Pagnamenta's model and the diamonds in the modified Mitra-Kumari's version. If one treats every nucleus just as a cluster of quarks, one may expect noticeable corrections up to nuclei with $A=10$.}
    \label{fig:quarkchangeradius}
\end{figure}

Although the results with light nuclei are encouraging, the method soon meets its limitation
due to the fact that the corrections in (\ref{nice}) and (\ref{MKradius}) are inversely proportional to the mass of the nucleus. With both the accuracy of measured radii
for heavy nuclei, that is, large nucleon number, $A$ and the correction also
becoming smaller, the method may not serve at the moment for heavy nuclei. 
In Figure (\ref{fig:quarkchangeradius}), we show the difference 
$\Delta<r>^2=<r^2>^0-<r^2>$, as a function of the nucleon number, $A$. 

\subsection{Mesons}
Having discussed the case of the proton and light nuclei, we consider the mesons as the next-best candidates to examine the correlation between the Lorentz boost correction of a
radius and the cluster structure of the system. 
Starting with the $\pi^+$ meson, there appears a problem related to the application of the formalisms. Due to the extremely light mass of the pion, the correction exceeds the value of the radius causing the corrected mean square radius to be negative.  A  limitation in the method is found for particles with masses lower than approximately one half of the mass of the proton. By analyzing step by step the multiple approximations made in \cite{LichtPagnamenta1} we propose that in the case of light hadrons, the motion relative to the center of mass of the constituents associated with the cluster cannot be neglected. Licht and Pagnamenta assume that the world lines of each component of the cluster are completely parallel to the world line of the center of mass. This assumption simplifies the calculation of the general Lorentz boost, but may be the origin of the limitation mentioned above. 

Finally, the other well studied meson is the $K^+$ which has a larger mass than the pion. As a meson, we model it as a cluster of a valence quark - anti-quark pair leading to the results in Table \ref{table:kaon} with $n$ = 2. In this case there is no experimental value from spectroscopy so we use a computed value in order to compare the expected value of the radius at rest. As can be seen, the corrected radius is closer to the calculated value using the LP formalism and the modified LP or Mitra-Kumari formalism overestimates the correction like in the case with the proton. The value shown here can be considered as a prediction 
to the one that may be obtained using spectroscopy methods in the planned experiments using kaonic atoms \cite{Kspec}.
\hfill

\hfill

\begin{figure}[H]
    \centering
    \begin{tikzpicture}[%
      x=1mm,%
    y=-1mm,%
    every node/.style={%
        shape=rectangle,
        anchor=north west,
        outer sep=0mm,
        draw=black},
    whitecell/.style={%
        fill=\whitecellcolor,
        minimum height=\cellheight mm,
        minimum width=\cellwidth mm},
    bigwhitecell/.style={%
        fill=\whitecellcolor,
        minimum height=\cellheight mm,
        minimum width=80 mm}]

    \node[whitecell] at (0,0){Formalism};
    \node[whitecell] at (40,0){$n=2$};
    \node[whitecell] at (0,7){Licht-Pagnamenta};
    \node[whitecell] at (40,7){\textbf{0.469}};
    \node[whitecell] at (0,14){Mitra-Kumari};
    \node[whitecell] at (40,14){0.317};
    \node[bigwhitecell] at (0,21){Experimental Reference values (fm)};
    \node[whitecell] at (0,28){Scattering};
    \node[whitecell] at (40,28){0.583(40)};
    \node[whitecell] at (0,35){Theory};
    \node[whitecell] at (40,35){0.490};

    \end{tikzpicture}

    \captionof{table}{Corrections to the Kaon nuclear charge radius in fm for different formalisms. The data from scattering \cite{KScattering} (measured $\sqrt{<r^2>}$) is from electron kaon collisions taken at the CERN SPS; the theoretical value of \cite{KTheory} $\sqrt{<r^2>^0}$ is calculated from Dyson-Schwinger equations in QCD.}
    \label{table:kaon}
\end{figure}

\section{Summary}\label{summary}

Form Factors (FF) are a fundamental tool to properly model aspects, characteristics and the structure of hadrons and nuclei. By applying a Lorentz boost transformation to the electromagnetic FF extracted from scattering experiments, it is shown that this procedure generates a relativistic correction to the charge radius of hadrons and nuclei. This formalism was once used to provide a possible solution to the Proton Radius Puzzle in a satisfactory and theoretically consistent way \cite{neelimalcp}. 

Here, we extend this approach to compare the light nuclear radii measured from spectroscopy (at rest) and scattering experiments by performing a Lorentz boost 
based on the formalism of Licht-Pagnamenta and Mitra-Kumari. The boost in both formalisms depends on the number, $n$, of substructures inside the object. As expected, boosting the radius obtained from scattering to the rest frame not only  brings it close to the measured spectroscopic value but also gives the clue for the best choice of the value $n$, the number of clusters inside the object.
In case of the proton, the obvious choice was $n$ = 3 (corresponding to the three valence quarks), however, the nucleus is a more complex strongly interacting system and the cluster structure inside a nucleus is not obvious. Finding the best value of $n$ which brings the spectroscopic value of the radius closest to the scattering one, we comment on the cluster structure of nuclei.

Within this procedure, the deuteron is found to be a six-quark cluster and $^3$He and $^4$He are found to be simply made of 3 and 4 nucleons respectively. 
It is appealing to continue the investigation with heavier nuclei. However, we face two limitations: (i) the correction to the radius by boosting it from a moving to the rest frame is inversely proportional to the mass of the object (see (\ref{nice}) and (\ref{MKradius}) and (ii) the data on scattering and spectroscopic determination of radii of heavier nuclei are scarce and not accurate.
Most of the scattering experiments were performed in the last century and the spectroscopy is done using regular atoms, in contrast to the precise muonic atom data for the proton, deuterium and helium.   

In the present work, we have pointed out a curious connection between the cluster structure of hadrons and nuclei and the Lorentz boost used to relate the measurements of their radii performed in different reference frames. We hope that the results of this work motivate new experiments on the charge radii of light nuclei beyond $A$ = 4, to investigate the clustering in nuclei. 
\begin{acknowledgments}
N.G.K. thanks the Faculty of Science, Universidad de Los Andes, Colombia, for financial support through Grant No. INV-2023-162-2841. 
\end{acknowledgments}

\bibliographystyle{naturemag} 
\bibliography{Referencias}

@article{Marek, 
    title={All electromagnetic form factors},
    volume={26},
    DOI={10.1088/0143-0807/26/4/001}, 
    number={4}, 
    journal={European Journal of Physics}, 
    author={Nowakowski, M. and Paschos, E. A. and Rodríguez, J. M.},
    year={2005}, 
    month=apr, 
    pages={545}
    }

@article{gao,
    author = {Gao, Haiyan},
    title = {NUCLEON ELECTROMAGNETIC FORM FACTORS},
    journal = {International Journal of Modern Physics E},
    volume = {12},
    number = {01},
    pages = {1-40},
    year = {2003},
    doi = {10.1142/S021830130300117X},
    URL = { https://doi.org/10.1142/S021830130300117X}
}

@article{Sachs62,
  title = {High-Energy Behavior of Nucleon Electromagnetic Form Factors},
  author = {Sachs, R. G.},
  journal = {Phys. Rev.},
  volume = {126},
  issue = {6},
  pages = {2256--2260},
  numpages = {0},
  year = {1962},
  month = {Jun},
  publisher = {American Physical Society},
  doi = {10.1103/PhysRev.126.2256},
  url = {https://link.aps.org/doi/10.1103/PhysRev.126.2256}
}

@article{dipoleff,
    title = {New Limit on the Electron Electric Dipole Moment},
    author = {Regan, B. C. and Commins, Eugene D. and Schmidt, Christian J. and DeMille, David},
    journal = {Phys. Rev. Lett.},
    volume = {88},
    issue = {7},
    pages = {071805},
    numpages = {4},
    year = {2002},
    month = {Feb},
    publisher = {American Physical Society},
    doi = {10.1103/PhysRevLett.88.071805},
    url = {https://link.aps.org/doi/10.1103/PhysRevLett.88.071805}
}

@article{fsetheory,
    title = {Theory of light hydrogenlike atoms},
    author = {Eides, M. I. and Grotch, H. and Shelyuto V. A.},
    journal = {Physics Reports},
    volume = {342},
    issue = {2-3},
    pages = {63--261},
    year = {2001},
    month = {Feb},
    doi = {https://doi.org/10.1016/S0370-1573(00)00077-6},
    url = {https://www.sciencedirect.com/science/article/pii/S0370157300000776}
}

@article{Corrected_cs,
    author = "Borah, Kaushik and Hill, Richard J. and Lee, Gabriel and Tomalak, Oleksandr",
    title = "{Parametrization and applications of the low-$Q^2$ nucleon vector form factors}",
    eprint = "2003.13640",
    archivePrefix = "arXiv",
    primaryClass = "hep-ph",
    reportNumber = "FERMILAB-PUB-20-124-T",
    doi = "10.1103/PhysRevD.102.074012",
    journal = "Phys. Rev. D",
    volume = "102",
    number = "7",
    pages = "074012",
    year = "2020"
}

@article{Corrected_decay,
    author = "Zeynali, K. and Bashiry, V. and Zolfagharpour, F.",
    title = "{Form factors and decay rate of B$_{c}^{*}$ $ \rightarrow$ D$_{s}$l$^{+}$l$^{-}$ decays in the QCD sum rules}",
    eprint = "1410.0526",
    archivePrefix = "arXiv",
    primaryClass = "hep-ph",
    doi = "10.1140/epja/i2014-14127-5",
    journal = "Eur. Phys. J. A",
    volume = "50",
    pages = "127",
    year = "2014"
}

@article{Corrected_el,
  title = {Lamb shift in the muonic deuterium atom},
  author = {Krutov, A. A. and Martynenko, A. P.},
  journal = {Phys. Rev. A},
  volume = {84},
  issue = {5},
  pages = {052514},
  numpages = {14},
  year = {2011},
  month = {Nov},
  publisher = {American Physical Society},
  doi = {10.1103/PhysRevA.84.052514},
  url = {https://link.aps.org/doi/10.1103/PhysRevA.84.052514}
}

@article{Sick_2018,
    title={Proton Charge Radius from Electron Scattering},
    volume={6},
    ISSN={2218-2004},
    url={https://www.mdpi.com/2218-2004/6/1/2},
    DOI={10.3390/atoms6010002},
    number={1},
    journal={Atoms},
    author={Sick, Ingo},
    year={2018}
}

@article{Stat1,
  title = {Threshold energies and poles for hadron physical problems by a model-independent universal algorithm},
  author = {Tripolt, R. -A. and Haritan, I. and Wambach, J. and Moiseyev, N.},
  journal = {Physics Letters B},
  volume = {774},
  pages = {411--416},
  year = {2017},
  month = {Nov},
  doi = {https://doi.org/10.1016/j.physletb.2017.10.001},
  url = {https://www.sciencedirect.com/science/article/pii/S0370269317308018}
}

@article{Stat2,
  title = {Nucleon-to-Roper electromagnetic transition form factors at large ${Q}^{2}$},
  author = {Chen, Chen and Lu, Ya and Binosi, Daniele and Roberts, Craig D. and Rodr\'{\i}guez-Quintero, Jose and Segovia, Jorge},
  journal = {Phys. Rev. D},
  volume = {99},
  issue = {3},
  pages = {034013},
  numpages = {13},
  year = {2019},
  month = {Feb},
  publisher = {American Physical Society},
  doi = {10.1103/PhysRevD.99.034013},
  url = {https://link.aps.org/doi/10.1103/PhysRevD.99.034013}
}

@article{codata2010,
  title = {CODATA recommended values of the fundamental physical constants: 2010},
  author = {Mohr, Peter J. and Taylor, Barry N. and Newell, David B.},
  journal = {Rev. Mod. Phys.},
  volume = {84},
  issue = {4},
  pages = {1527--1605},
  numpages = {0},
  year = {2012},
  month = {Nov},
  publisher = {American Physical Society},
  doi = {10.1103/RevModPhys.84.1527},
  url = {https://link.aps.org/doi/10.1103/RevModPhys.84.1527}
}

@article{Exp1,
  author = {Aldo Antognini  and François Nez  and Karsten Schuhmann  and Fernando D. Amaro  and François Biraben  and João M. R. Cardoso  and Daniel S. Covita  and Andreas Dax  and Satish Dhawan  and Marc Diepold  and Luis M. P. Fernandes  and Adolf Giesen  and Andrea L. Gouvea  and Thomas Graf  and Theodor W. Hänsch  and Paul Indelicato  and Lucile Julien  and Cheng-Yang Kao  and Paul Knowles  and Franz Kottmann  and Eric-Olivier Le Bigot  and Yi-Wei Liu  and José A. M. Lopes  and Livia Ludhova  and Cristina M. B. Monteiro  and Françoise Mulhauser  and Tobias Nebel  and Paul Rabinowitz  and Joaquim M. F. dos Santos  and Lukas A. Schaller  and Catherine Schwob  and David Taqqu  and João F. C. A. Veloso  and Jan Vogelsang  and Randolf Pohl },
  title = {Proton Structure from the Measurement of 2S-2P Transition Frequencies of Muonic Hydrogen},
  journal = {Science},
  volume = {339},
  number = {6118},
  pages = {417-420},
  year = {2013},
  doi = {10.1126/science.1230016},
  URL = {https://www.science.org/doi/abs/10.1126/science.1230016},
  eprint = {https://www.science.org/doi/pdf/10.1126/science.1230016}
}

@article{Exp2,
    author = {Axel Beyer  and Lothar Maisenbacher  and Arthur Matveev  and Randolf Pohl  and Ksenia Khabarova  and Alexey Grinin  and Tobias Lamour  and Dylan C. Yost  and Theodor W. Hänsch  and Nikolai Kolachevsky  and Thomas Udem },
    title = {The Rydberg constant and proton size from atomic hydrogen},
    journal = {Science},
    volume = {358},
    number = {6359},
    pages = {79-85},
    year = {2017},
    doi = {10.1126/science.aah6677},
    URL = {https://www.science.org/doi/abs/10.1126/science.aah6677},
    eprint = {https://www.science.org/doi/pdf/10.1126/science.aah6677},
}

@article{Exp3,
  title = {New Measurement of the $1S\ensuremath{-}3S$ Transition Frequency of Hydrogen: Contribution to the Proton Charge Radius Puzzle},
  author = {Fleurbaey, H\'el\`ene and Galtier, Sandrine and Thomas, Simon and Bonnaud, Marie and Julien, Lucile and Biraben, Fran\ifmmode \mbox{\c{c}}\else \c{c}\fi{}ois and Nez, Fran\ifmmode \mbox{\c{c}}\else \c{c}\fi{}ois and Abgrall, Michel and Gu\'ena, Jocelyne},
  journal = {Phys. Rev. Lett.},
  volume = {120},
  issue = {18},
  pages = {183001},
  numpages = {5},
  year = {2018},
  month = {May},
  publisher = {American Physical Society},
  doi = {10.1103/PhysRevLett.120.183001},
  url = {https://link.aps.org/doi/10.1103/PhysRevLett.120.183001}
}

@article{Exp4,
    author = {N. Bezginov  and T. Valdez  and M. Horbatsch  and A. Marsman  and A. C. Vutha  and E. A. Hessels },
    title = {A measurement of the atomic hydrogen Lamb shift and the proton charge radius},
    journal = {Science},
    volume = {365},
    number = {6457},
    pages = {1007-1012},
    year = {2019},
    doi = {10.1126/science.aau7807},
    URL = {https://www.science.org/doi/abs/10.1126/science.aau7807},
    eprint = {https://www.science.org/doi/pdf/10.1126/science.aau7807}
}

@article{Exp5,
    author = {Cui, Zhu-Fang and Binosi, Daniele and Roberts, Craig D. and Schmidt, Sebastian M.},
    title = {Fresh Extraction of the Proton Charge Radius from Electron Scattering},
    journal = {Phys. Rev. Lett.},
    volume = {127},
    number = {9},
    pages = {092001},
    year = {2021},
    doi = {10.1103/PhysRevLett.127.092001},
    URL = {https://link.aps.org/doi/10.1103/PhysRevLett.127.092001},
    eprint = {}
}

@article{Miller2019,
  title = {Confinement in Nuclei and the Expanding Proton},
  author = {Miller, Gerald A.},
  journal = {Phys. Rev. Lett.},
  volume = {123},
  issue = {23},
  pages = {232003},
  numpages = {6},
  year = {2019},
  month = {Dec},
  publisher = {American Physical Society},
  doi = {10.1103/PhysRevLett.123.232003},
  url = {https://link.aps.org/doi/10.1103/PhysRevLett.123.232003}
}

@article{LinHammer2022,
  title = {New Insights into the Nucleon's Electromagnetic Structure},
  author = { Y-H. Lin and H-W. Hammer and U-G. Meissner },
  journal = {Phys. Rev. Lett.},
  volume = {128},
  issue = {9},
  pages = {052002},
  numpages = {5},
  year = {2022},
  month = {Feb},
  publisher = {American Physical Society},
  doi = {10.1103/PhysRevLett.128.052002},
  url = {https://doi.org/10.1103/PhysRevLett.128.052002}
}

@article{Lorenz2012,
  title = {The size of the proton: Closing in on the radius puzzle},
  author = { I. T. Lorenz and H-W. Hammer and U-G. Meissner },
  journal = {Eur. Phys. J. A},
  volume = {48},
  issue = {11},
  pages = {151},
  numpages = {5},
  year = {2012},
  month = {Nov},
  publisher = {Springer Nature},
  doi = {10.1140/epja/i2012-12151-1},
  url = {https://doi.org/10.1140/epja/i2012-12151-1}
}

@article{neelimalcp,
    title={Lorentz contracted proton},
    volume={2015},
    ISSN={1029-8479},
    DOI={10.1007/JHEP09(2015)215},
    abstractNote={The proton charge and magnetization density distributions can be related to the well known Sachs electromagnetic form factors GE,M(q2) through Fourier transforms, only in the Breit frame. The Breit frame however moves with relativistic velocities in the Lab and a Lorentz boost must be applied before extracting the static properties of the proton from the corresponding densities. Apart from this, the Fourier transform relating the densities and form factors is inherently a non-relativistic expression. We show that the relativistic corrections to it can be obtained by extending the standard Breit equation to higher orders in its 1/c2 expansion. We find that the inclusion of the above corrections reduces the size of the proton as determined from electron proton scattering data by about 4%.},
    number={9},
    journal={Journal of High Energy Physics},
    author={Bedoya Fierro, D. and Kelkar, N. G. and Nowakowski, M.},
    year={2015},
    month=sep,
    pages={215}
}

@article{PRAD,
    author={Xiong, W. and Gasparian, A. and Gao, H. and Dutta, D. and Khandaker, M. and Liyanage, N. and Pasyuk, E. and Peng, C. and Bai, X. and Ye, L. and Gnanvo, K. and Gu, C. and Levillain, M. and Yan, X. and Higinbotham, D. W. and Meziane, M. and Ye, Z. and Adhikari, K. and Aljawrneh, B. and Bhatt, H. and Bhetuwal, D. and Brock, J. and Burkert, V. and Carlin, C. and Deur, A. and Di, D. and Dunne, J. and Ekanayaka, P. and El-Fassi, L. and Emmich, B. and Gan, L. and Glamazdin, O. and Kabir, M. L. and Karki, A. and Keith, C. and Kowalski, S. and Lagerquist, V. and Larin, I. and Liu, T. and Liyanage, A. and Maxwell, J. and Meekins, D. and Nazeer, S. J. and Nelyubin, V. and Nguyen, H. and Pedroni, R. and Perdrisat, C. and Pierce, J. and Punjabi, V. and Shabestari, M. and Shahinyan, A. and Silwal, R. and Stepanyan, S. and Subedi, A. and Tarasov, V. V. and Ton, N. and Zhang, Y. and Zhao, Z. W.},
    title={A small proton charge radius from an electron--proton scattering experiment},
    journal={Nature},
    year={2019},
    month={Nov},
    day={01},
    volume={575},
    number={7781},
    pages={147-150},
    issn={1476-4687},
    doi={10.1038/s41586-019-1721-2},
    url={https://doi.org/10.1038/s41586-019-1721-2}
}

@article{jentschura2022,
    author ={Jentschura, U. D.},
    title = {Proton Radius: A Puzzle or a Solution!?},
    journal = {Journal of Physics: Conference Series},
    year = {2022},
    month={Mar},
    day={22},
    volume={2391},
    pages={012017},
    doi={10.1088/1742-6596/2391/1/012017},
    url={https://iopscience.iop.org/article/10.1088/1742-6596/2391/1/012017/pdf}
}

@article{lumpay2025,
    author = "Lumpay, Roland B. and Jusoy, Jade C. and Apas, Ruel and Auxtero, Eulogio",
    title = "{The Proton Radius Puzzle and Discrepancies in Proton Structure Measurements}",
    eprint = "2501.11195",
    archivePrefix = "arXiv",
    primaryClass = "nucl-ex",
    month = "1",
    year = "2025"
}

@article{qi2025,
  title = {Toward resolving the discrepancy in helium-3 and helium-4 nuclear charge radii},
  author = {Qi, Xiao-Qiu and Zhang, Pei-Pei and Yan, Zong-Chao and Tang, Li-Yan and Chen, Ai-Xi and Shi, Ting-Yun and Zhong, Zhen-Xiang},
  journal = {Phys. Rev. Res.},
  volume = {7},
  issue = {2},
  pages = {L022020},
  numpages = {6},
  year = {2025},
  month = {Apr},
  publisher = {American Physical Society},
  doi = {10.1103/PhysRevResearch.7.L022020},
  url = {https://link.aps.org/doi/10.1103/PhysRevResearch.7.L022020}
}

@article{HillEPJweb2017,
    author ={Hill, R. J.},
    title = {Review of experimental and theoretical status of the proton radius puzzle},
    journal = {EPJ Web of Conferences},
    year = {2017},
    month={Mar},
    day={22},
    volume={137},
    doi={10.1051/epjconf/201713701023},
    url={https://www.epj-conferences.org/articles/epjconf/abs/2017/06/epjconf_conf2017_01023/epjconf_conf2017_01023.html}
}

@article{Gao2022,
  title = {The proton charge radius},
  author = {Gao, H. and Vanderhaeghen, M.},
  journal = {Rev. Mod. Phys.},
  volume = {94},
  issue = {1},
  pages = {015002},
  numpages = {43},
  year = {2022},
  month = {Jan},
  publisher = {American Physical Society},
  doi = {10.1103/RevModPhys.94.015002},
  url = {https://link.aps.org/doi/10.1103/RevModPhys.94.015002}
}

@article{LichtPagnamenta1,
  title = {Wave Functions and Form Factors for Relativistic Composite Particles. I},
  author = {Licht, Arthur Lewis and Pagnamenta, Antonio},
  journal = {Phys. Rev. D},
  volume = {2},
  issue = {6},
  pages = {1150--1156},
  numpages = {0},
  year = {1970},
  month = {Sep},
  publisher = {American Physical Society},
  doi = {10.1103/PhysRevD.2.1150},
  url = {https://link.aps.org/doi/10.1103/PhysRevD.2.1150}
}

@article{MitraKumari,
  title = {Relativistic form factors for clusters with nonrelativistic wave functions},
  author = {Mitra, A. N. and Kumari, Indra},
  journal = {Phys. Rev. D},
  volume = {15},
  issue = {1},
  pages = {261--266},
  numpages = {0},
  year = {1977},
  month = {Jan},
  publisher = {American Physical Society},
  doi = {10.1103/PhysRevD.15.261},
  url = {https://link.aps.org/doi/10.1103/PhysRevD.15.261}
}

@article{asymptotic,
  title = {Asymptotic form factors of hadrons and nuclei and the continuity of particle and nuclear dynamics},
  author = {Brodsky, Stanley J. and Chertok, Benson T.},
  journal = {Phys. Rev. D},
  volume = {14},
  issue = {11},
  pages = {3003--3020},
  numpages = {0},
  year = {1976},
  month = {Dec},
  publisher = {American Physical Society},
  doi = {10.1103/PhysRevD.14.3003},
  url = {https://link.aps.org/doi/10.1103/PhysRevD.14.3003}
}

@article{WeiNST2024,
  title = {Clustering in nuclei: progress and perspectives},
  author = {Wei, K. and Ye, Y-L. and Yang, Zai-Hong},
  journal = {Nuclear Science and Techniques},
  volume = {35},
  issue = {12},
  year = {2024},
  month = {Nov},
  doi = {10.1007/s41365-024-01588-x},
  url = {https://doi.org/10.1007/s41365-024-01588-x}
}

@article{HoyleApJS1954,
  title = {On Nuclear Reactions Occuring in Very Hot STARS.I. the Synthesis of Elements from Carbon to Nickel},
  author = {{Hoyle}, F.},
  journal = {apjs},
  volume = {1},
  pages = {121},
  year = {1954},
  month = {Sep},
  doi = {10.1086/190005},
  url = {https://ui.adsabs.harvard.edu/abs/1954ApJS....1..121H/abstract}
}

@article{FreerRPP2007,
    doi = {10.1088/0034-4885/70/12/R03},
    url = {https://dx.doi.org/10.1088/0034-4885/70/12/R03},
    year = {2007},
    month = {nov},
    publisher = {},
    volume = {70},
    number = {12},
    pages = {2149},
    author = {Freer, Martin},
    title = {The clustered nucleus—cluster structures in stable and unstable nuclei},
    journal = {Reports on Progress in Physics}
}

@article{BuckPRC1995,
  title = {Unified treatment of scattering and cluster structure in \ensuremath{\alpha}+closed shell nuclei: $^{20}\mathrm{Ne}$ and $^{44}\mathrm{Ti}$},
  author = {Buck, B. and Johnston, J. C. and Merchant, A. C. and Perez, S. M.},
  journal = {Phys. Rev. C},
  volume = {52},
  issue = {4},
  pages = {1840--1844},
  year = {1995},
  month = {Oct},
  publisher = {American Physical Society},
  doi = {10.1103/PhysRevC.52.1840},
  url = {https://link.aps.org/doi/10.1103/PhysRevC.52.1840}
}

@article{BagchiEPJA2020,
  title = {Signature of a possible $\alpha$ - cluster state in $N$ = $Z$ doubly-magic $^{56}$ Ni},
  author = {Bagchi, S. and Akimune, H. and Gibelin, J. and Harakeh, M. N. and Kalantar-Nayestanaki, N. and Achouri, N. L. and Bastin, B. and Boretzky, K. and Bouzomita, H. and Caama{\~{n}}o, M. and C{\`a}ceres, L. and Damoy, S. and Delaunay, F. and Fern{\'a}ndez-Dom{\'i}nguez, B. and Fujiwara, M. and Garg, U. and Grinyer, G. F. and Kamalou, O. and Khan, E. and Krasznahorkay, A. and Lhoutellier, G. and Libin, J. F. and Lukyanov, S. and Mazurek, K. and Najafi, M. A. and Pancin, J. and Penionzhkevich, Y. and Perrot, L. and Raabe, R. and Rigollet, C. and Roger, T. and Sambi, S. and Savajols, H. and Senoville, M. and Stodel, C. and Suen, L. and Thomas, J. C. and Vandebrouck, M. and Walle, J. Van de}, journal={The European Physical Journal A},
  year={2020},
  month={Nov},
  day={12},
  volume={56},
  number={11},
  pages={290},
  issn={1434-601X},
  doi={10.1140/epja/s10050-020-00294-7},
  url={https://doi.org/10.1140/epja/s10050-020-00294-7}
}

@article{FranzGrossEPJC2023,
  title = {50 Years of quantum chromodynamics},
  author={Gross, Franz and Klempt, Eberhard and Brodsky, Stanley J. and Buras, Andrzej J. and Burkert, Volker D. and Heinrich, Gudrun and Jakobs, Karl and Meyer, Curtis A. and Orginos, Kostas and Strickland, Michael and Stachel, Johanna and Zanderighi, Giulia and Brambilla, Nora and Braun-Munzinger, Peter and Britzger, Daniel and Capstick, Simon and Cohen, Tom and Crede, Volker and Constantinou, Martha and Davies, Christine and Del Debbio, Luigi and Denig, Achim and DeTar, Carleton and Deur, Alexandre and Dokshitzer, Yuri and Dosch, Hans G{\"u}nter and Dudek, Jozef and Dunford, Monica and Epelbaum, Evgeny and Escobedo, Miguel A. and Fritzsch, Harald and Fukushima, Kenji and Gambino, Paolo and Gillberg, Dag and Gottlieb, Steven and Grafstrom, Per and Grazzini, Massimiliano and Grube, Boris and Guskov, Alexey and Iijima, Toru and Ji, Xiangdong and Karsch, Frithjof and Kluth, Stefan and Kogut, John B. and Krauss, Frank and Kumano, Shunzo and Leinweber, Derek and Leutwyler, Heinrich and Li, Hai-Bo and Li, Yang and Malaescu, Bogdan and Mariotti, Chiara and Maris, Pieter and Marzani, Simone and Melnitchouk, Wally and Messchendorp, Johan and Meyer, Harvey and Mitchell, Ryan Edward and Mondal, Chandan and Nerling, Frank and Neubert, Sebastian and Pappagallo, Marco and Pastore, Saori and Pel{\'a}ez, Jos{\'e} R. and Puckett, Andrew and Qiu, Jianwei and Rabbertz, Klaus and Ramos, Alberto and Rossi, Patrizia and Rustamov, Anar and Sch{\"a}fer, Andreas and Scherer, Stefan and Schindler, Matthias and Schramm, Steven and Shifman, Mikhail and Shuryak, Edward and Sj{\"o}strand, Torbj{\"o}rn and Sterman, George and Stewart, Iain W. and Stroth, Joachim and Swanson, Eric and de T{\'e}ramond, Guy F. and Thoma, Ulrike and Vairo, Antonio and van Dyk, Danny and Vary, James and Virto, Javier and Vos, Marcel and Weiss, Christian and Wobisch, Markus and Wu, Sau Lan and Young, Christopher and Yuan, Feng and Zhao, Xingbo and Zhou, Xiaorong},
  journal={The European Physical Journal C},
  year={2023},
  month={Dec},
  day={12},
  volume={83},
  number={12},
  pages={1125},
  issn={1434-6052},
  doi={10.1140/epjc/s10052-023-11949-2},
  url={https://doi.org/10.1140/epjc/s10052-023-11949-2}
}

@article{tsaiPTP1980,
  title = {Multi-Quark States: Their Classification, Production and Possible Presence in Nuclei},
  author = {Tsai, S. Y.},
  journal = {Prog. Theor. Phys.},
  volume = {64},
  pages = {1710},
  year = {1980},
  doi = {10.1143/PTP.64.1710},
  url = {https://doi.org/10.1143/PTP.64.1710}
}

@article{millerPRL1984,
  title = {Six-Quark Cluster Components of Nuclear Wave Functions and the Pion-Nucleus Double - Charge - Exchange Reaction},
  author = {Miller, Gerald A.},
  journal = {Phys. Rev. Lett.},
  volume = {53},
  issue = {21},
  pages = {2008--2011},
  numpages = {0},
  year = {1984},
  month = {Nov},
  publisher = {American Physical Society},
  doi = {10.1103/PhysRevLett.53.2008},
  url = {https://link.aps.org/doi/10.1103/PhysRevLett.53.2008}
}

@article{kochmillerPRC1985,
  title = {Six quark cluster effects and binding energy differences between mirror nuclei},
  author = {Koch, Volker and Miller, Gerald A.},
  journal = {Phys. Rev. C},
  volume = {31},
  issue = {2},
  pages = {602--612},
  numpages = {0},
  year = {1985},
  month = {Feb},
  publisher = {American Physical Society},
  doi = {10.1103/PhysRevC.31.602},
  url = {https://link.aps.org/doi/10.1103/PhysRevC.31.602}
}

@article{kankiPTP1985,
    title = {Formation of Six-Quark Clusters in Nuclei and EMC Effect},
    author = {Kanki, Takeshi and Miyamura, Osamu},
    journal = {Progress of Theoretical Physics},
    volume = {73},
    number = {2},
    pages = {414-420},
    year = {1985},
    month = {02},
    issn = {0033-068X},
    doi = {10.1143/PTP.73.414},
    url = {https://doi.org/10.1143/PTP.73.414},
    eprint = {https://academic.oup.com/ptp/article-pdf/73/2/414/5231459/73-2-414.pdf},
}

@article{glozmanPPNP1995,
  author = "Glozman, L. Ya.",
    editor = "Faessler, Amand",
    title = "{The six-quark structure of the deuteron and reactions D(e,e' p)Delta,N*}",
    doi = "10.1016/0146-6410(95)00009-8",
    journal = "Prog. Part. Nucl. Phys.",
    volume = "34",
    pages = "123--132",
    year = "1995"
}

@article{burovZPA1982,
  title = {On the six-quark structure in the deuteron form factor},
  author={Burov, V. V. and Dorkin, S. M. and Lukyanov, V. K. and Titov, A. I.},
  journal={Zeitschrift f{\"u}r Physik A Atoms and Nuclei},
  year={1982},
  month={Jun},
  day={01},
  volume={306},
  number={2},
  pages={149-154},
  issn={0939-7922},
  doi={10.1007/BF01415484},
  url={https://doi.org/10.1007/BF01415484}
}

@article{codata2018,
  title = {CODATA recommended values of the fundamental physical constants: 2018},
  author = {Tiesinga, Eite and Mohr, Peter J. and Newell, David B. and Taylor, Barry N.},
  journal = {Rev. Mod. Phys.},
  volume = {93},
  issue = {2},
  pages = {025010},
  numpages = {63},
  year = {2021},
  month = {Jun},
  publisher = {American Physical Society},
  doi = {10.1103/RevModPhys.93.025010},
  url = {https://link.aps.org/doi/10.1103/RevModPhys.93.025010}
}

@article{CREMAH,
    title={The size of the proton},
    volume={466}, ISSN={1476-4687},
    DOI={10.1038/nature09250},
    number={7303},
    journal={Nature}, author={Pohl, Randolf and Antognini, Aldo and Nez, François and Amaro, Fernando D. and Biraben, François and Cardoso, João M. R. and Covita, Daniel S. and Dax, Andreas and Dhawan, Satish and Fernandes, Luis M. P. and Giesen, Adolf and Graf, Thomas and Hänsch, Theodor W. and Indelicato, Paul and Julien, Lucile and Kao, Cheng-Yang and Knowles, Paul and Le Bigot, Eric-Olivier and Liu, Yi-Wei and Lopes, José A. M. and Ludhova, Livia and Monteiro, Cristina M. B. and Mulhauser, Françoise and Nebel, Tobias and Rabinowitz, Paul and dos Santos, Joaquim M. F. and Schaller, Lukas A. and Schuhmann, Karsten and Schwob, Catherine and Taqqu, David and Veloso, João F. C. A. and Kottmann, Franz},
    year={2010},
    month=jul,
    pages={213–216}
}

@article{magneticprot,
  title = {Precise determination of the proton magnetic radius from electron scattering data},
  author = {Alarc\'on, J. M. and Higinbotham, D. W. and Weiss, C.},
  journal = {Phys. Rev. C},
  volume = {102},
  issue = {3},
  pages = {035203},
  numpages = {7},
  year = {2020},
  month = {Sep},
  publisher = {American Physical Society},
  doi = {10.1103/PhysRevC.102.035203},
  url = {https://link.aps.org/doi/10.1103/PhysRevC.102.035203}
}

@article{specmagprot,
  title = {Model-independent determination of the magnetic radius of the proton from spectroscopy of ordinary and muonic hydrogen},
  author = {Karshenboim, Savely G.},
  journal = {Phys. Rev. D},
  volume = {90},
  issue = {5},
  pages = {053013},
  numpages = {9},
  year = {2014},
  month = {Sep},
  publisher = {American Physical Society},
  doi = {10.1103/PhysRevD.90.053013},
  url = {https://link.aps.org/doi/10.1103/PhysRevD.90.053013}
}

@article{truemagprot,
    title={Zemach and magnetic radius of the proton from the hyperfine splitting in hydrogen},
    volume={33},
    ISSN={1434-6079},
    DOI={10.1140/epjd/e2005-00025-9},
    number={1},
    journal={The European Physical Journal D - Atomic, Molecular, Optical and Plasma Physics},
    author={Volotka, A. V. and Shabaev, V. M. and Plunien, G. and Soff, G.},
    year={2005},
    month=apr,
    pages={23–27}
}

@article{6quarkexp,
    title={A signal of the six-quark cluster in deuteron electrodisintegration},
    volume={386},
    ISSN={0370-2693},
    DOI={https://doi.org/10.1016/0370-2693(96)00901-X},
    number={1},
    journal={Physics Letters B},
    author={Lu, Lan-Chun and Cheng, Tan-Sheng},
    year={1996},
    pages={69–74}
}

@mastersthesis{6quarks,
  author  = "Nunes, A. M.",
  title   = "The deuteron as a six-quark state in QCD",
  school  = "Universidade de Lisboa",
  year    = "2023"
}

@article{codata2014,
    title={CODATA Recommended Values of the Fundamental Physical Constants: 2014*}, volume={45},
    ISSN={0047-2689},
    DOI={10.1063/1.4954402},
    number={4},
    journal={Journal of Physical and Chemical Reference Data},
    author={Mohr, Peter J. and Newell, David B. and Taylor, Barry N.},
    year={2016},
    month=nov,
    pages={043102}
}

@article{CREMAD,
    title={Laser spectroscopy of muonic deuterium},
    volume={353},
    DOI={10.1126/science.aaf2468},
    number={6300},
    journal={Science},
    author={Pohl, Randolf and Nez, François and Fernandes, Luis M. and Amaro, Fernando D. and Biraben, François and Cardoso, João M. and Covita, Daniel S. and Dax, Andreas and Dhawan, Satish and Diepold, Marc and et al.},
    year={2016},
    month={Aug},
    pages={669–673}
}

@article{maizePRC1985,
  title = {Multi-quark compound states and the $^{3}\mathrm{He}$ charge form factor},
  author = {Maize, M. A. and Kim, Y. E.},
  journal = {Phys. Rev. C},
  volume = {31},
  issue = {5},
  pages = {1923--1928},
  numpages = {0},
  year = {1985},
  month = {May},
  publisher = {American Physical Society},
  doi = {10.1103/PhysRevC.31.1923},
  url = {https://link.aps.org/doi/10.1103/PhysRevC.31.1923}
}

@article{angeli2013,
  title = {Table of experimental nuclear ground state charge radii: An update},
  author = {Angeli, I. and Marinova, K. P.},
  journal = {Atomic Data and Nuclear Data Tables},
  volume = {99},
  issue = {1},
  pages = {69-95},
  numpages = {},
  year = {2013},
  month = {Jan 2013},
  publisher = {},
  doi = {10.1016/j.adt.2011.12.006},
  url = {https://doi.org/10.1016/j.adt.2011.12.006}
}

@article{jaffe1968,
  title = {Binding energies and charge radii of $^3\mathrm{H}$ and 
$^3\mathrm{He}$},
  author = {Jaffe, A. I. and Reiner, A. S.},
  journal = {Phys. Lett. B},
  volume = {26},
  issue = {12},
  pages = {719-720},
  numpages = {},
  year = {1968},
  month = {May 1968},
  publisher = {},
  doi = {10.1016/0370-2693(68)90401-2},
  url = {https://doi.org/10.1016/0370-2693(68)90401-2}
}

@article{folk1968,
  title = {Coulomb energy of $^3\mathrm{H}$ and $^3\mathrm{He}$},
  author = {Folk, R. T.},
  journal = {Phys. Lett. B},
  volume = {28},
  issue = {},
  pages = {159-160},
  numpages = {},
  year = {1968},
  month = {Oct 1968},
  publisher = {},
  doi = {10.1016/0370-2693(68)90002-6},
  url = {https://doi.org/10.1016/0370-2693(68)90002-6}
}

@article{vanasse2017,
  title = {Triton charge radius to next-to-next-to-leading order in pionless effective field theory},
  author = {Vanasse, Jared},
  journal = {Phys. Rev. C},
  volume = {95},
  issue = {2},
  pages = {024002},
  numpages = {25},
  year = {2017},
  month = {Feb},
  publisher = {American Physical Society},
  doi = {10.1103/PhysRevC.95.024002},
  url = {https://link.aps.org/doi/10.1103/PhysRevC.95.024002}
}

@article{Myers2016,
  title = {The $^3\mathrm{H}$-$^3\mathrm{He}$ charge radii difference},
  author = {Myers, L. S. and Arrington, J. R. and Higinbotham, D. W.},
  journal = {EPJ Web Conf.},
  volume = {113},
  issue = {},
  pages = {08013},
  numpages = {2},
  year = {2016},
  month = {},
  publisher = {},
  doi = {10.1051/epjconf/201611308013},
  url = {https://doi.org/10.1051/epjconf/201611308013}
}

@article{pohltalk,
  title = {Towards a Cold Atomic Hydrogen Source at $\mathrm{T-REX}$ $\mathrm{M}$ainz },
  author = {Haack, J. and Heppener, M. and Schmidt, S. and Schumacher, H-L. and Willig, M. and Wieltsch, A. and Pohl, R.},
  journal = {},
  volume = {},
  issue = {},
  pages = {},
  numpages = {},
  year = {2019},
  month = {},
  publisher = {},
  doi = {},
  url = {https://www.agpohl.physik.uni-mainz.de/files/2019/03/Plakat_DPG19_Jan_Haack_v3.pdf}
}

@article{he3scattering,
    title = {Zemach moments of $^{3}\mathrm{He}$ and $^{4}\mathrm{He}$},
    author = {Sick, Ingo},
    journal = {Phys. Rev. C},
    volume = {90},
    issue = {6},
    pages = {064002},
    numpages = {3},
    year = {2014},
    month = {Dec},
    publisher = {American Physical Society},
    doi = {10.1103/PhysRevC.90.064002},
    url = {https://link.aps.org/doi/10.1103/PhysRevC.90.064002}
}

@article{he3muon,
    author = "Schuhmann, Karsten and others",
    collaboration = "CREMA",
    title = "{The helion charge radius from laser spectroscopy of muonic helium-3 ions}",
    eprint = "2305.11679",
    archivePrefix = "arXiv",
    primaryClass = "physics.atom-ph",
    month = "5",
    year = "2023"
}

@article{he4scattering,
  title = {Precise root-mean-square radius of $^{4}\mathrm{He}$},
  author = {Sick, Ingo},
  journal = {Phys. Rev. C},
  volume = {77},
  issue = {4},
  pages = {041302},
  numpages = {3},
  year = {2008},
  month = {Apr},
  publisher = {American Physical Society},
  doi = {10.1103/PhysRevC.77.041302},
  url = {https://link.aps.org/doi/10.1103/PhysRevC.77.041302}
}

@article{he4muon,
    title={Measuring the $\alpha$-particle charge radius with muonic helium-4 ions},
    volume={589},
    ISSN={1476-4687},
    DOI={10.1038/s41586-021-03183-1},
    number={7843},
    journal={Nature},
    author={Krauth, Julian J. and Schuhmann, Karsten and Ahmed, Marwan Abdou and Amaro, Fernando D. and Amaro, Pedro and Biraben, François and Chen, Tzu-Ling and Covita, Daniel S. and Dax, Andreas J. and Diepold, Marc and Fernandes, Luis M. P. and Franke, Beatrice and Galtier, Sandrine and Gouvea, Andrea L. and Götzfried, Johannes and Graf, Thomas and Hänsch, Theodor W. and Hartmann, Jens and Hildebrandt, Malte and Indelicato, Paul and Julien, Lucile and Kirch, Klaus and Knecht, Andreas and Liu, Yi-Wei and Machado, Jorge and Monteiro, Cristina M. B. and Mulhauser, Françoise and Naar, Boris and Nebel, Tobias and Nez, François and dos Santos, Joaquim M. F. and Santos, José Paulo and Szabo, Csilla I. and Taqqu, David and Veloso, João F. C. A. and Vogelsang, Jan and Voss, Andreas and Weichelt, Birgit and Pohl, Randolf and Antognini, Aldo and Kottmann, Franz},
    year={2021},
    month=jan,
    pages={527–531}
}

@article{16o1,
    author = "Bertsch, G. F. and Bertozzi, W.",
    title = "{\ensuremath{\alpha}-Particle model of 16 O}",
    doi = "10.1016/0375-9474(71)90158-8",
    journal = "Nucl. Phys. A",
    volume = "165",
    pages = "199--210",
    year = "1971"
}

@article{16o2,
    author = "Tohsaki, A. and Horiuchi, H. and Schuck, P. and Ropke, G.",
    title = "{Alpha cluster condensation in C-12 and O-16}",
    eprint = "nucl-th/0110014",
    archivePrefix = "arXiv",
    doi = "10.1103/PhysRevLett.87.192501",
    journal = "Phys. Rev. Lett.",
    volume = "87",
    pages = "192501",
    year = "2001"
}

@article{16o3,
    author = "Ohkubo, S. and Hirabayashi, Y. and Hirabayashi, Y.",
    title = "{alpha-particle condensate states in O-16}",
    eprint = "1102.1762",
    archivePrefix = "arXiv",
    primaryClass = "nucl-th",
    doi = "10.1016/j.physletb.2009.12.066",
    journal = "Phys. Lett. B",
    volume = "684",
    pages = "127--131",
    year = "2010"
}

@article{Kspec,
    author = {Michel, Niklas and Oreshkina, Natalia S.},
    title = {Access to the Kaon Radius with Kaonic Atoms},
    journal = {Annalen der Physik},
    volume = {534},
    number = {3},
    pages = {2100150},
    doi = {https://doi.org/10.1002/andp.202100150},
    url = {https://onlinelibrary.wiley.com/doi/abs/10.1002/andp.202100150},
    eprint = {https://onlinelibrary.wiley.com/doi/pdf/10.1002/andp.202100150},
    year = {2022}
}

@article{KScattering,
    title={A measurement of the kaon charge radius},
    volume={178},
    ISSN={0370-2693},
    DOI={https://doi.org/10.1016/0370-2693(86)91407-3},
    number={4},
    journal={Physics Letters B},
    author={Amendolia, S. R. and Batignani, G. and Beck, G. A. and Bellamy, E. H. and Bertolucci, E. and Bologna, G. and Bosisio, L. and Bradaschia, C. and Budinich, M. and Dell’orso, M. and Piazzoli, B. D’Ettorre and Fabbri, F. L. and Fidecaro, F. and Foa, L. and Focardi, E. and Frank, S. G. F. and Gianetti, P. and Giazzotto, A. and Giorgi, M. A. and Green, M. G. and Heath, G. P. and Landon, M. P. J. and Laurelli, P. and Liello, F. and Mannocchi, G. and March, P. V. and Marrocchesi, P. S. and Menzione, A. and Meroni, E. and Picchi, P. and Ragusa, F. and Ristori, L. and Rolandi, L. and Scribano, A. and Stefanini, A. and Storey, D. and Strong, J. A. and Tenchini, R. and Tonelli, G. and Triggiani, G. and Schlippe, W. von and Zallo, A.},
    year={1986},
    pages={435–440}
}

@article{KTheory,
    title={Electromagnetic form factors of charged and neutral kaons},
    volume={371},
    ISSN={0370-2693},
    DOI={https://doi.org/10.1016/0370-2693(96)00006-8},
    number={3}, journal={Physics Letters B},
    author={Burden, C. J. and Roberts, C. D. and Thomson, M. J.},
    year={1996},
    pages={163–168}
}
\end{document}